# On Dark Energy and Dark Matter
# (Part I)

Shlomo Barak and Elia M Leibowitz
School of Physics & Astronomy, Tel Aviv University
arxiv:astro-ph v1     24 December 2008

## Abstract

Phenomena currently attributed to Dark Energy (DE) and Dark Matter (DM) are merely a result of the interplay between gravitational energy density, $\epsilon_g$, generated by the contraction of space by matter, and the energy density of the Cosmological Microwave Background (CMB), $\epsilon_{CMB}$ which causes space dilation.

These contentions and the observation that, globally, in the universe $\epsilon_g \approx \epsilon_{CMB}$ lead to the derivation of the Hubble parameter, H, as a function of the scale factor, a, the time, t, the redshift, z, and to the calculation of its present value, $H_0$. They also lead to a new understanding of the cosmological redshift and the Euclidian nature of the universe.

From H(t) we conclude that $\dot{a} = \text{const}$, ($\ddot{a} = 0$), in contrast to the consensus of the last decade. This result is supported by the fit of our theoretically derived flux from supernovae (SN) as a function of z, with observation. This flux is derived based on our H(z) that determines $D_L$, the Luminosity Distance (LD). We obtain this fit without any free parameters, whereas in current cosmology this fit is obtained by using the dependent free parameters $\Omega_M$ and $\Omega_\Lambda$.

---

This paper is divided into three parts. Part I (this paper) discusses DE and Cosmological issues. Part II (arxiv:astro-ph 2008) addresses the DM issue, where we show that in and around galaxies the above interplay causes inhomogeneous and anisotropic expansion of space. The expansion of the universe is homogeneous and isotropic only when viewed globally. This leads to a theoretical derivation of dynamic and kinematic relations that fit observed Rotation Curves (RC) in galaxies. Part III introduces a model that describes the creation and evolution of galaxies and their resulting RCs.

The ideas and notions raised in this trilogy are more fully discussed in a forthcoming book by one of us, S. Barak (2009) that reconsiders the foundations of Physics.

## Dark Energy

DE has been suggested to explain the **supposed** changes in the rate of expansion of the universe, and to explain its Euclidian nature (flatness), A. G. Riess et al (1998), S. Perlmutter et al (1999). However, we show that $\dot{a} = \text{const}$, ($\ddot{a} = 0$).

To prove our contentions, with which we resolve the issues of DE and DM, we reconsider basic concepts in cosmology and astrophysics. Assumptions presented below are then justified theoretically, and are shown to be valid by observations.

1. **Space is three-dimensional, foamy, elastic and vibrating.**
   The **assumption** that space is 3D, deformed or un-deformed, means that the universe is **not** a curved 3D manifold in a hyperspace with an additional spatial dimension. For a **flat universe** the issue of an additional spatial dimension is not relevant. The terms "deformed" and "curved" are used for a 3D-space and a 3D-manifold, respectively.

   The consensus that space is foamy, and hence cellular, rests on the meaning of expansion, and the requirement that its vibrations have a finite energy density.

   The deformation of space is the change in size of its cells. Positive or negative deformation, around a point in space, means that the space cells grow or shrink, respectively, from this point outwards. For a positively curved manifold, the ratio of the circumference of a circle to the radius is less than $2\pi$, as measured by a yardstick of fixed length. For a deformed 3D-space, with a positive deformation, around a point the above ratio is also less than $2\pi$, as measured by a flexible yardstick such as the linear dimension of a space cell. Note that for a **deformed space** there is no meaning to global deformation, **deformation is a local attribute**. The surface of a sphere with radius R is a 2D manifold with a global curvature 1/R. However, a global deformation for a deformable 2D planar sheet can only have the value zero i.e., the sheet is Euclidian. The smallest linear dimension of a space cell, whether Planck's length or not, which determines the Zero Point Fluctuations (ZPF) energy density, is not relevant to our discussion. Here, deformation around a point is expressed by a scale factor a(r,t) that depends on both time and the vector, r, from the point. Space density is determined by a(r,t).

   We suggest that space vibrations are the Electromagnetic (EM) waves and that these waves dilate space (in solids it is anharmonicity that is responsible for thermal expansion). Hence, globally, the CMB is the main contributor to space dilation over the average space density set by the ZPF, A. D. Sakharov (1968) and C. W. Misner et al (1970). We thus consider the CMB to be a **global** anti-gravitational **cosmological constant**, $\Lambda = \frac{8\pi G}{c^4} \cdot \in_{CMB}$, see Section 3. We show that the present value of $\in_{CMB}$ determines the present value, $H_0$, of the Hubble parameter.

   Part II shows that **locally** around a mass the interplay between $\in_g$ and $\in_{CMB}$ causes an inhomogeneous expansion of space, which enhances positive deformation around the mass. This explains the "DM" phenomena. We also show that $\in_{CMB}$ at the time of formation of a galaxy determines, $g_0$, the central acceleration at which its Rotation Curve (RC) becomes flat. $g_0 = 1.2 \cdot 10^{-8}$ cm sec$^{-2}$ for galaxies formed approximately 12 BY ago, the epoch of galaxy formation.

   $\in = mc^2$ does not imply that all forms of energy deform space in the same way. We contend that energy in the form of matter deforms space positively by contracting it, whereas energy in the form of electromagnetic waves deforms space by dilating it. The above has **implications** for both the mass equivalence principle and the way that the energy densities of gravitation and electromagnetic waves are incorporated in the General Relativity (GR) field equation. This is discussed in a book by S. Barak (2009).

2. **Space contraction or expansion is the change of its cell size.**
   GR shows that a mass **contracts** space around it. Cells close to the mass are smaller than those at a distance and hence the elastic **positive** deformation of space. Length, close to a mass, is smaller, and the "running of time" is slower, than at a distance. Gravitation is the elastic deformation of space, remove the mass and the deformation is gone. In Part II we discuss the issue of elastic versus non-elastic 3D space. Deformation due to space expansion is non-elastic, as "DM halos" show.

   Space expansion is the enlargement of its cells, and the CMB dilating vibrational energy contributes to this expansion. This implies that the number of space cells in the universe is conserved.

   Space contraction by a mass can be represented by a gravitational scale factor $a_g \equiv \ell/\ell_0$, where $\ell_0$ is a distance in space far from masses, and $\ell$ is the same distance contracted by the introduction of a mass. In GR, $\ell/\ell_0 = \exp(\varphi/c^2)$, hence $a_g$ at the surface of the sun, or at the edge of our galaxy, is approximately $1 - 10^{-6}$ whereas in the last 12 BY the scale factor used in cosmology changed, due to expansion, from 0.25 to 1.
   This difference, orders of magnitude, in space deformation is related to the elastic versus non-elastic behavior of space, see Part II.

3. **The contribution of electromagnetic energy density to space deformation must be reconsidered.**

   The dilation of space by the energy density of Electromagnetic (EM) waves, such as the CMB, is a non-linear phenomenon, which implies the violation of Lorentz Invariance (LI).

   Examples of other non-linear EM phenomena are the bending of light beams in a gravitational field, and the scattering of intense light by light which results in the production of pairs of electrons and positrons. No matter is present in the space occupied by the interacting beams (D. L. Burk et al, 1997 and G. Brodin et al, 2002). Maxwell's equations, however, being linear, express only linear EM phenomena.

   This paper provides proof of our assertion that the CMB dilates space. Part II presents an additional proof – the theoretical derivation of both the dynamic and kinematic laws of RCs that resemble the Milgrom and Tully Fisher relations. This derivation dispels the mystery of DM.

4. **The cosmological principle (CP) implies different geometries of the universe for different types of space.**
   For a 3D space manifold, curved in a hyperspace with an extra spatial dimension, CP implies a uniform global curvature. In this case the universe is finite but with no boundary. For a 3D space CP implies flatness. This is the result of curving being a local attribute only. In other words the only global curvature possible is zero. In this case the universe is either infinite, or finite with a boundary. Close to a boundary CP can not hold true.
   The measured energy density of the CMB, as of today, is $\epsilon_{CMB_0} \sim 4 \cdot 10^{-13}\ \mathrm{erg\,cm^{-3}}$, whereas the calculated energy density of baryonic matter is:
   $\epsilon_m = mc^2 \sim 2 \cdot 10^{-10}\ \mathrm{erg\,cm^{-3}}$. This estimation is based on the Big Bang Nucleosynthesis

(BBN) range of values for m, and is **independent** of the Hubble constant, W. Rindler (2004) p.337.

Thus the universe is mass dominant, but **in contrast** to current understanding, this does not imply that the interior of the universe must be positively curved.

5. **The contractual gravitational energy of the universe equals its dilational CMB energy.**
Gravitational energy is confined in each and every Hubble Sphere (HS) since gravitational contraction moves at the speed of light relative to the mass that is generating the gravitational field. Every point in the universe is both the center and the edge of some identical HSs. Therefore, $\epsilon_g$, far from masses, must be, globally, the same. This is used as an argument in our explanation of the Euclidian nature of the universe.

$$\epsilon_g = \frac{1}{8\pi G} E_g^2 = \frac{1}{8\pi G}\left(\frac{GM}{R_{HS}^2}\right)^2 \quad \text{hence:}$$

(1) $\quad \epsilon_g = \frac{2\pi G}{9} m^2 R_{HS}^2 = \frac{2\pi G}{9} m^2 \frac{c^2}{H^2}$

m – baryonic mass density of the universe   $R_{HS}$ – radius of an HS.

The present calculated value of $\epsilon_g$ taking m = $2 \cdot 10^{-31}$ gm cm$^{-3}$ based on BBN is:
$\epsilon_g = 3.5 \cdot 10^{-13}$ erg cm$^{-3}$
whereas the present measured value for the $\epsilon_{CMB}$ is:
$\epsilon_{CMB} = 4.17 \cdot 10^{-13}$ erg cm$^{-3}$
This indicates that $\langle \epsilon_g \rangle = \langle \epsilon_{CMB} \rangle$ or, equivalently that the gravitational energy $U_g$ equals the CMB energy $U_{CMB}$ in the universe. Note that, whereas $\epsilon_g$ is concentrated around masses, $\epsilon_{CMB}$ is distributed homogeneously. $\epsilon_g$ and $\epsilon_{CMB}$ depend on $a^{-4}$, hence their equality is retained over time.

That $U_g$ equals $U_{CMB}$ is not accidental. Elementary particles are considered point-like or string-like and structureless, but if they are not (S. Barak, 2009) then the above equality is explained.

Inside a finite size elementary particle ZPF waves with a wavelength twice its diameter cannot resonate. Hence, in the process of creation of the particle, energy emigrates from the inside to the outside – Casimir effect. The emigrated energy is above the ZPF and hence is detectable – this is the CMB energy. The deficiency in energy inside causes contraction around which is gravitation, with the same contractual gravitational energy as the energy that left the particle.

Note that our suggestion for the origin of the CMB explains its uniformity in the universe **without** the need for a supposed epoch of inflation. This energy, through its interaction with matter, acquired its blackbody characteristics.
We thus conclude that regardless of the structure of an elementary particle and just as a result of its being of finite size, $U_g = U_{CMB}$. This understanding leads to the proof of the mass equivalence principle (S, Barak, 2009).

Our conviction that this is the case arises from the fact that this equality leads to the derivation of H(a) and the calculation of $H_0$, the Hubble constant, with an accuracy of ~15%, see below. We do not know of any theory that can yield the value of $H_0$.
The expansion of the universe is the result $\epsilon_{CMB}$ being homogeneous whereas $\epsilon_g$ is concentrated around galaxies. In the inter-galactic space $\epsilon_{CMB} \gtrsim \epsilon_g$ hence the expansion ($\epsilon$ is pressure) and in and around galaxies $\epsilon_g > \epsilon_{CMB}$ and hence expansion is inhibited. Part II shows that this inhomogeneous expansion results in the flattening of Rotation Curves (RC).
Inhabitants of a 3D universe can only make observations related to internal deformations. However, such deformations, on a global scale, do not appear since $U_g$ and $U_{CMB}$ are distributed, on a large scale, homogeneously. The result is the Euclidian nature of the universe, and hence the validity of CP far from the boundary.

In no way is a **critical mass density** or the **idea of inflation**, involved in our considerations and calculations.

6. **We derive the dependence (evolution) of the Hubble parameter, H, on the scale factor, a, by equating $\epsilon_{CMB}$ to the gravitational energy density $\epsilon_g$ in the universe and calculate its present value, $H_0$.**
From equation (1):
$$\epsilon_g = \frac{2\pi G}{9} m^2 \frac{c^2}{H^2} = \epsilon_{CMB}$$
m – baryonic mass density of the universe    $R_{HS}$ – radius of an HS.
Values as of today are designated by the suffix 0. Note that the scale factor, a, as of today is chosen as 1 and hence in the past was less than 1. The above equation gives:

(2) $\qquad H^2 = \dfrac{2\pi G c^2 m^2}{9 \epsilon_{CMB}}$

which is our "equivalent" to the FRW result.
Substituting $m = m_0 \cdot a^{-3}$ and the known relation, $\epsilon_{CMB} = \epsilon_{CMB_0} \cdot a^{-4}$ in Equation (2) gives:

$$H^2 = \frac{2\pi G}{9} \frac{m_0^2 \cdot c^2}{\epsilon_{CMB_0}} \cdot a^{-2} \qquad \text{hence:}$$

$$H = m_0 \cdot c \sqrt{\frac{2\pi G}{9 \epsilon_{CMB_0}}} \cdot a^{-1} \qquad \text{and thus:}$$

(3) $\qquad H = H_0 \, a^{-1} \qquad$ where:

(4) $\quad H_0 = m_0 \cdot c \sqrt{\dfrac{2\pi G}{9\,\epsilon_{CMB_0}}}\quad$ Inserting the measured values:

$c = 3 \cdot 10^{10}$ cm sec$^{-1}$ $\quad \epsilon_{CMB_0} = 4.17 \cdot 10^{-13}$ erg cm$^{-3}$,
and the BBN estimation $m_0 = 2 \cdot 10^{-31}$ gr cm$^{-3}$ in the above equation gives:
$H_0 = 2 \cdot 10^{-18}$ sec$^{-1}$.

The measured value, $H_0 = (72 \pm 8)$ km sec$^{-1}$/Mpc $= (2.3 \pm 0.26) \cdot 10^{-18}$ sec$^{-1}$ (W. L. Freedman et al, 2000) is obtained with the higher estimation, $2.3 \cdot 10^{-31}$ gm cm$^{-3}$, for $m_0$.

7. **We derive the Hubble Parameter, H as a function of time, t, and show that $\dot{a} = $ const ($\ddot{a} = 0$).**
This result is in contrast to the consensus but as Section 9 shows, it complies with observation.

$H \stackrel{def}{=} \dot{a}/a$, but, as we have shown $H = H_0/a$, hence:
$\dot{a} = $ const ($\ddot{a} = 0$).

The value of the constant $\dot{a}$ is $2.3 \cdot 10^{-18}$ sec$^{-1}$ since $H_0 = \dot{a}/a_0$ where $a_0 = 1$ today
Integrating both sides of $da = H_0 dt$ and designating BB - Big Bang, $_0$ – now, $t_{BB} = 0$, $a_0 = 1$, gives:

$a - a_{BB} = H_0 t \quad$ but: $\quad a = \dfrac{H_0}{H} \quad$ hence:

(5) $\quad t = \dfrac{1}{H} - \dfrac{1}{H_{BB}} \quad\quad$ Now, at $t = t_0$, $H = H_0$ hence:

(6) $\quad t_0 = \dfrac{1}{H_0} - \dfrac{1}{H_{BB}} \quad\quad$ Since $H_{BB} \gg H_0$

$t_0 \sim \dfrac{1}{H_0} = 13.7$ BY $\quad$ **the age of the universe**

(7) $\quad H(t) = \dfrac{1}{t + 1/H_{BB}}$

The distance between any two galaxies grows with a, but H falls with a. We thus conclude that any two galaxies recede from each other at all times, at a constant velocity $v = r \cdot H$. This has implications for our understanding of the Big Bang.

8. **Cosmological redshift due to space expansion is $z = e^{\frac{v_r}{c}} - 1$**
The basis for our discussion is:

- A yardstick and a clock, far from masses are **not** affected by their location in space or by time.

- $d = \ell a$, the distance between two points in an expanding universe, as measured by a yardstick.
  a is the scale factor with its present chosen value $a = 1$ hence in the past $a < 1$.

$\ell$ is defined as the ratio $d/a$. Alternatively, $d = na$ where n is the number of space cells and a is the linear dimension of a cell.

- Light velocity, c, is a constant of nature, affected only by the presence of mass.
- The distance between a photon, or crest of a wave, and its emitter is: $d(t) = \ell(t)a(t)$ and its velocity relative to the emitter (located in galaxy A) is:

(8) $\quad v_p = \dot{d}(t) = \dot{\ell}(t) \cdot a(t) + \ell \cdot \dot{a}(t)$

The first term on the right-hand side is light velocity:

(9) $\quad c = \dot{\ell} a$

whereas the second term is the recessional velocity of the place at which the photon is momentary "present".

$v_r = \ell \dot{a}(t) \quad$ therefore:

$v_p = c + v_r$

In this discussion, Special Relativity is not relevant.

We have shown that $\ddot{a} = \text{const}$ hence:

(10) $\quad a(t) = a(t_z) + \dot{a}(t - t_z) = a_z + \dot{a}(t - t_z) \quad$ for any time $t \geq t_z$.

In our discussion we use $t_z$ as the time of emission of our photon from galaxy A.

From $c = \dot{\ell} a$ we get:

$$\frac{d\ell}{dt} = \frac{c}{a} = \frac{c}{a_z + \dot{a}(t - t_z)}$$

$$\frac{1}{c} d\ell = \frac{dt}{a_z + \dot{a}(t - t_z)} \quad \text{and by integration:}$$

$$\frac{1}{c}(\ell - \ell_z) = \frac{1}{\dot{a}}\left\{\ln[a_z + \dot{a}(t - t_z)] - \ln a_z\right\} \quad \text{but} \quad \ell_z = 0, \text{ hence:}$$

$$\frac{\ell}{c} = \frac{1}{\dot{a}} \cdot \ln\left[1 + \frac{\dot{a}}{a_z}(t - t_z)\right] \quad\quad \text{but} \quad \frac{\dot{a}}{a_z} = H(t_z) = H_z, \quad \text{hence:}$$

$$t - t_z = \frac{1}{H_z}\left[\exp\left(\frac{\dot{a}\ell}{c}\right) - 1\right]$$

Let $t_0$ be the cosmic time of arrival of a wavecrest to the observer in galaxy B.

$t - t_z = \frac{1}{H_z}\left[\exp\left(\frac{\dot{a}\ell(t_0)}{c}\right) - 1\right] \quad$ but $\quad \ell(t_0) = \ell_0 = d_0 \quad$ is the present distance between galaxies A and B. Hence $\dot{a}\ell_0$ is their recessional velocity $v_r$, and thus the Look Back Time is:

(11) $\quad (t - t_z) = \frac{1}{H_z}\left[e^{\frac{v_r}{c}} - 1\right]$

The cosmological redshift is the result of successive crests arriving at the observer with a longer arrival time difference $\Delta t'$ than their time difference $\Delta t$ at emission.

$\lambda_{em} = c \cdot \Delta t$  whereas  $\lambda_{obs} = c \cdot \Delta t'$.

$\Delta t' = t_2 - t_1 \quad \Delta t = t_{z_2} - t_{z_1}$

$\Delta t' = \left[(t_2 - t_{z_2}) - (t_1 - t_{z_1})\right] + \Delta t$

Equation (11) gives:

(12) $\Delta t' = \left[\dfrac{1}{H(t_2)} - \dfrac{1}{H(t_1)}\right] \cdot \left[e^{\frac{v}{c}} - 1\right] + \Delta t$   and since:

(13) $H(t) = \dfrac{\dot{a}}{a} = \dfrac{\dot{a}}{a_z + \dot{a}(t - t_z)} = \dfrac{\frac{\dot{a}}{a_z}}{1 + \frac{\dot{a}}{a_z}(t - t_z)} = \dfrac{H}{1 + H_z(t - t_z)}$

our z as a function of the recessional velocity is:

(14) $z = \dfrac{\lambda_{obs}}{\lambda_{em}} - 1 = e^{\frac{v_r}{c}} - 1$

For $v_r = c$ we get $z = 1.718$.

GR, using $\Omega_M = 0.3$ and $\Omega_\Lambda = 0.7$, gives for $v_r = c$, the value $z \gtrsim 1.5$. See Fig. 2 in

T. M. Davis and C. H. Lineweaver (2003). This has implications for the Particle Horizon and the other cosmological issues as discussed in the referenced paper.

We derive the known relation, $a = 1/(1+z)$, from equation (9).

For dt the time difference between successive crests:

$c = \dfrac{d\ell}{dt} \cdot a$      gives:

$\lambda_{obs} = c \cdot dt_{obs} = d\ell \cdot a_{obs}$

$\lambda_{em} = c \cdot dt_{em} = d\ell \cdot a_{em}$    hence:

$\dfrac{\lambda_{obs}}{\lambda_{em}} = \dfrac{a_{obs}}{a_{em}}$    For $a_{obs} = 1$ and $a_{em} \equiv a$:

$\dfrac{\lambda_{obs}}{\lambda_{em}} = \dfrac{1}{a}$

$z = \dfrac{\lambda_{obs}}{\lambda_{em}} - 1 = \dfrac{1}{a} - 1$   which gives:

$a = \dfrac{1}{1+z}$

This is the result of differences in the arrival times of successive wave crests, and is not related to the concept of a photon.

This result is valid only for the dynamic case of an expanding universe.

9. **The relations $H = H_0 a^{-1}$ and $a = 1/(1+z)$ give $H(z) = H_0(1 + z)$. We show that this result is confirmed by observations.**

   In this section we use the conventional notation $H(z) = H_0 h(z)$. In **our** theory:

   (15)    $h(z) = 1 + z$

   Whereas the **known** equation with the two dependent free parameters, $\Omega_M$ and $\Omega_\Lambda$, for **flat space** where $\Omega_M + \Omega_\Lambda = 1$ is:

   (16)    $h(z) = \left[ \Omega_M (1+z)^3 + \Omega_\Lambda \right]^{\frac{1}{2}}$

   see A. G. Riess et al (2006) (2007) and S. M. Carroll (2004).

   Our $h(z)$ yields a different $d_L$, the Luminosity Distance (LD) from that derived from equation (16). LD is defined by the ratio of the luminosity, L, of a supernova, to its measured flux, F:

   (17)    $d_L^2 \equiv \dfrac{L}{4\pi F}$

   From the known relation:

   (18)    $d_L = (1+z) \dfrac{c}{H_0} \cdot \int_0^z \dfrac{dz'}{h(z')}$     using **our** $h(z)$, we get:

   (19)    $d_L = (1+z) \dfrac{c}{H_0} \cdot \int_0^z \dfrac{dz'}{1+z'} = (1+z) \dfrac{c}{H_0} \ln(1+z)$

   Equation (13) is the basis of the following comparison of our theory with observations. Figure (1) plots the **distance modulus** µ versus the log of the **redshifts**, z, with data from 307 supernovae Ia (blue squares).

   Note that while the LD derived in FRW cosmology is a function with two dependent free parameters, $\Omega_M + \Omega_\Lambda = 1$ (flat universe) that can be, and were in fact adjusted to various data sets, in the last decade, our LD is obtained directly from theory, with **no free parameters**.

   Figure (1) shows the distance modulus versus log(z), where z is the cosmological redshift.

   **Frame (a)** displays the measured data in 307 SN Ia as compiled and presented in the website of the SNP Union  http://supernova.lbl.gov/Union, Kowalski et al (2008).
   The three curves are, from bottom up:
   (i) (Green) derived from the FRW cosmology for a flat and matter-dominated universe with $(\Omega_M, \Omega_\Lambda) = (1.0, 0.0)$.
   (ii) (Red) derived by our cosmology with no acceleration.
   (iii) (Black) derived from the FRW cosmology for an accelerating flat universe with $(\Omega_M, \Omega_\Lambda) = (0.3, 0.7)$.

   **Frame (b)** zoom on the upper-right corner of Frame (a), shows the fit of the three curves to the data.

**Frame (c)** extends the theoretical curves to $z = 100$. The family of thin lines are FRW lines for 11 pairs of the flat-universe parameters starting from $\Omega_M = 0.10$, $\Omega_\Lambda = 1 - \Omega_M$ in steps of 0.1. The three heavy lines are the same as in Frames (a) and (b). The highest thin line is that of an FRW $(\Omega_M, \Omega_\Lambda) = (0.0, 1.0)$ which represent a dark energy-dominated universe.

**Frame (d)** is a zoom on the upper-right corner of Frame (c). It shows the theoretical curves for large z values.
Note the cross-over between our curve (Red) and the curve (Black) of the currently accepted $\Lambda$ CMB cosmology that occurs around $\log(z) = 0.76$, $z = 5.76$. Thus for higher z values our cosmology predicts a distance modulus as a function of $\log(z)$ curve that is distinctly different from that predicted by $\Lambda$ CMB cosmology for a flat universe. Future observations of high z standard candles should support our cosmology.

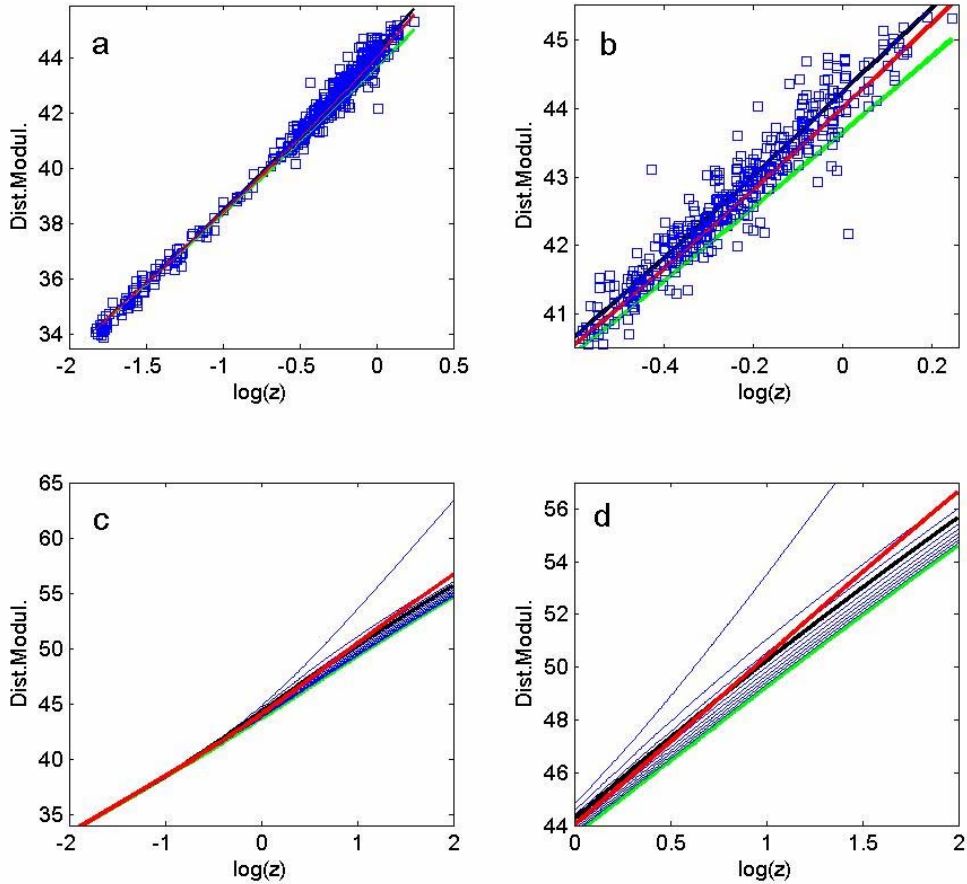

**Figure (1) Distance Modulus, μ, Versus the Log of the Redshift, z
with Data Points for 307 Ia Supernovae**

## Discussion

It is now appropriate to relate to the apparent failure of the current cosmology.

We agree that GR is valid.

We agree that the FRW metric is correct:
The velocity of light is a constant of nature, independent of space density.
Our result, $\dot{a} = \text{const}$, enables us to consider the expansion of the universe to be the uni-directional flow of time. Thus the expansion is a master clock, which started with the BB, having a unit of time determined by rate of expansion.

However, despite the above, we find that Friedmann's equations are irrelevant to cosmology. This is the result of $\epsilon_g$ being equal to $\epsilon_{CMB}$ which affects the GR field equation and renders it irrelevant to cosmology. This issue will be discussed in full elsewhere.

## Summary

In the universe the gravitational energy equals the CMB energy and hence $\epsilon_g \approx \epsilon_{CMB}$. This implies that $\dot{a} = \text{const}$, ($\ddot{a} = 0$).

The above leads to $H(z) = H_0(1+z)$ and hence to $D_L \propto \ln(1 + z)$. This result is supported by its fit to data from observations of hundreds of Ia supernovae - obtained without any free parameters. This validates our theoretical result that $\dot{a} = \text{const}$.

The Hubble constant, $H_0$, is calculated and the result fits the calculated value based on observations.

Our reconsideration of basic concepts in Cosmology and Astrophysics show that $\epsilon_{CMB}$ and $\epsilon_g$ are sufficient to account for cosmological phenomena, and thus dispel the mystery of Dark Energy.

## Acknowledgements

We would like to thank Roger M. Kaye for his linguistic contribution and technical assistance.